\documentstyle[aps,prb,epsf]{revtex}
\begin{document}

\title{Zeeman splitting of shallow donors in GaN}

\author{
 Francisco Mireles and Sergio E. Ulloa
       }

\address {
 Department of Physics and Astronomy, and Condensed Matter and
 Surface Sciences Program \\ Ohio University, Athens, Ohio 45701--2979
         }

\date{Jun 01 1998}

\twocolumn

\maketitle

 \begin{abstract}

The Zeeman splitting of the donor spectra in cubic- and hexagonal-GaN
are studied using an effective mass theory approach. Soft-core
pseudopotentials were used to describe the chemical shift of the
different substitutional dopants. The donor ground states calculated
range from 29.5 to 33.7 meV, with typically 1 meV higher binding in the
hexagonal phase. Carbon is found to produce the largest donor binding
energy. The ionization levels and excited states are in excellent
agreement with Hall and optical measurements, and suggest the presence
of residual C in recent experiments.

 \end{abstract}


\narrowtext

Shallow donors in III-N semiconductors are one of the key fundamental
elements in the fabrication of short-wave length optical and
high-frequency electronic devices.\cite{Strite} Despite the great
activity and considerable progress on the growth quality and
characterization of these materials, the origin of the n-type
conductivity in nominally undoped GaN is still not well established.
Native defects, such as N vacancies, and residual impurities like O and
Si have been invoked as the prime sources of the high electron
concentrations ($\sim 10^{17}\, cm^{-3}$) seen in
experiments.\cite{Perlin,Gotz} However, the donor ionization
levels introduced are not fully known. Temperature dependent transport
experiments have indicated binding energies of about 36 meV for
residual donors and 26 meV for Si. \cite{Gaskill} Recent Hall effect
measurements for high-energy electron-irradiated GaN/sapphire samples
suggest the appearance of a donor state from the N vacancy with a donor
energy of about 64 meV, much larger than the value expected for
residual donors.\cite{Look} Study of electronic excitations by Raman
scattering of unintentional donors gives energies ranging from 18.5 to
30 meV for cubic and hexagonal GaN. \cite{Ramsteiner} Although O is
known to display a shallow level state, it behaves as a deep donor
under high hydrostatic pressure ($>20$ GPa).  However, this is not the
case for Si.\cite{Wetzel}

Arrhenius plot analyses of the electron concentration of O- and Si-
implanted in GaN yields activation energies of about 29 meV for both
dopants. \cite{Gotz} On the other hand, magneto-transmission
measurements of Zeeman splitting of the donor spectra in Si-doped GaN
films are a bit higher, and consistent with other reports. \cite{Wang}
However, more recent infrared spectroscopy studies of
the Zeeman splitting of donors in undoped GaN grown by hydride vapor
phase epitaxy (HVPE) show $1s-2p_{\pm}$ transitions at 23.3 and 25.95
meV (in the limit of zero magnetic field), apparently due to two
different donor species.\cite{Moore} Ground states at 31.1 and 33.8
meV, respectively, were inferred from these transitions. While the
lower transition was believed to come from Si, the origin of the larger
one was unclear.

Carbon is known to be an amphoteric impurity in GaN, but it is widely
believed to predominantly introduce acceptor levels. Recent quantum
molecular dynamics calculations by Boguslawski {\it et al.}
\cite{Bogus} showed that C substituting in the Ga site gives rise to
shallow donor states. Although the formation energy estimates indicated
that C atoms do prefer to occupy the N site, there is still the
possibility that some C atoms will be incorporated into the Ga
vacancies. In these calculations, C was found to have also an excellent
solubility, and an efficient self-compensation of acceptors by carbon
donors.\cite{Bogus} Moreover, Zhang {\it et al.} \cite{Zhang} have
recently reported photoluminescence experiments of C-doped GaN which
clearly demonstrate that the incorporation of carbon can produce a
significant yellow luminescence, and that there is a continued
C-compensation by the generation of additional shallow donors.

In this work we calculate binding energies for substitutional donor
impurities for different species. Our results indicate that residual
carbon in GaN films could indeed give rise to additional shallow donors
with binding energy close to those previously assigned to Si and O. To
our knowledge, the essential effect of central cell corrections for the
different impurities on the donor ionization levels in GaN has not been
discussed before.  We also introduce here the effective mass anisotropy
in wurtzite GaN, which is shown to be important in the calculation of
impurity states.

The purpose of this letter is then twofold. First, we report effective
mass theory (EMT) calculations of the binding energy of Si, O and C
shallow donors in GaN for both crystal phases, cubic (zincblende) and
hexagonal (wurtzite). In this EMT calculations, the anisotropy of the
conduction band and the dielectric constant for h-GaN are explicitly
included. Furthermore, the central cell corrections introduced by the
different impurities are taken into account via soft-core atomic
pseudopotentials.  The second purpose is to examine the Zeeman
splitting of the excited states of substitutional impurities under the
presence of an external magnetic field, and to compare our estimates
with the available experimental results.

Near the $\Gamma$ point ($k=0$), the lower conduction band \cite{LCB}
in hexagonal GaN is ellipsoidal, with different longitudinal and
transverse (with the c-axis) effective masses (which are clearly equal
in the cubic phase). Within effective mass theory, and neglecting spin
splitting, \cite{spin} the Hamiltonian for the envelope wave function
under an external magnetic field along the $\hat z$-axis (parallel to
the crystal c-axis), is given by
\begin{eqnarray}
H & = &-\frac{\hbar ^2}{2m_{\perp }^{*}}\left( \frac{\partial
^2}{\partial \,x^2}+\frac{\partial ^2}{\partial
\,y^2} \right) -\frac{\hbar ^2 }{2m_{\Vert }^{*}} \frac{\partial
^2}{\partial \,z^2}+ \mu_{\perp }^{*} L_z B_z  \nonumber \\
   & + & \frac{\pi \mu_{\perp}^* B_z^2}{2 \Phi _o}(x^2+y^2)+U(r)\, ,
\end{eqnarray}
 where we have used a symmetric gauge, $\mu_{\perp}^*$ is the effective
Bohr magneton, $\mu_{\perp}^*=e\hbar /2m_{\perp}^*c$, and $L_z$ is the
orbital momentum operator along the principal axis (in units of
$\hbar$). Here, $\Phi _o= hc/e$ denotes the magnetic flux quantum, and
$U(r)$ is the impurity-donor potential defined as the difference
between the bare pseudopotential of the impurity and that of the host
atom.\cite{Mireles} The pseudopotentials we use here are those of
Jansen and Sankey, \cite{Jansen} who fitted to the functional form of
Bachelet-Hamann-Schl\"uter fame. \cite{Bachelet} These pseudopotentials
consist essentially of two parts. A local short-range
angular-momentum-dependent potential $V_l(r)$, and a long-range core
potential $V_{core}(r)$, where
 \begin{equation}
V_l(r)=\sum\limits_{i=1}^3[A_i(l)+r^2A_{i+3}(l)]\,e^{-\alpha _i(l)\,r^2}
\, ,
\end{equation}
 and
\begin{equation}
V_{core}(r)=-\frac{Z_v}r\sum\limits_{i=1}^2C_i\,{\rm erf}[(\alpha
_i^{core})^{1/2}r]
\, .
\end{equation}
 Here, $A_i(l),\, \alpha _i(l), \, C_i$ and $\alpha _i^{core}$ are the
fitting parameters to {\it ab initio} pseudopotentials, with $Z_v$
denoting the valence charge. This type of pseudopotential has been used
widely and with great success in calculations of the electronic and
structural properties of semiconductors. \cite{Jansen} The
extremely good transferability of these pseudopotentials in the
description of atoms in different chemical environments has been
discussed before by Goedecker {\it et al.}, \cite{Goe} and is related
to the existence of an inert region around the nucleus, essentially
independent of the chemical environment.

The envelope eigenfunctions $F(r)$ of the Hamiltonian in Eq.\ (1) are
obtained in a variational approach, and chosen here to be an expansion
of spherical harmonics and a linear combination of hydrogenic radial
functions such that $F(r)=\sum B_j\,r^le^{-\beta_jr}\,Y_{lm}(\theta
,\phi )$. The donor energy levels are then obtained using the sets
$\{B_j \}$ and $\{\beta _j\}$ of variational parameters.

For c-GaN (zincblende), we have used a reliable theoretical value of
the isotropic effective mass for the conduction band
($m_{c}^*=0.19\,m_{o}$),\cite{kim} while for the dielectric constant we
have used the experimental (and isotropic) value of $\epsilon_o$= 9.5.
\cite{Strite} [Notice that $m_c^*$ is not known experimentally.] On the
other hand, for the hexagonal (wurtzite) phase of GaN we have used the
experimental (and theoretical \cite{kim}) value of
$m_{\perp}^*=0.22\,m_{o}$, \cite{Moore,Perlin1} and the theoretical
value $m_{\Vert}^*=0.19\,m_{o}$,\cite{kim} for the parallel mass. In
this case the dielectric constant is obtained from the combination
$\epsilon_{o}^{*}=(2\epsilon_{\perp}^{-1}/3 +
\epsilon_{\Vert}^{-1}/3)^{-1} = 9.8$, where $\epsilon_{\perp}$ and
$\epsilon_{\Vert}$ are set to the experimental values of 9.5 and 10.4,
respectively.\cite{Madelung} Due to the moderately polar behavior of
these materials, the effect of the electron-phonon interaction on the
shift of the energy levels is considered through a polaron correction.
A Fr\"ohlich coupling constant of $\alpha_F=0.49$ has been fixed for
both crystal phases. This value gives rise to an 8\% enhancement of the
donor states. \cite{Sak}

Figure 1 shows a plot of the position of the electronic transitions
$1s-2p_{o}$ and $1s-2p_{\pm}$ as a function of magnetic field
$\boldmath B$ for O and C donors in h-GaN. The field is parallel to the
c-axis direction. The ground state and transitions to the first excited
states at zero field are shown in Table I.\@ For O we find a ground
state energy of 31.4 meV, and 33.7 for C, with internal transition
energies $1s-2p_{\pm}$ at 23.3 meV and 25.1 meV respectively. These
values are in excellent agreement to those found by Moore {\it et al.}
\cite{Moore} at 23.3 and 25.9 meV. Notice we have used the same
effective mass ($m_{\perp}^*=0.22\,m_{o}$) and dielectric constant
($\epsilon_{o}^*=9.8$) as those authors. However, they assumed that the
lower transition arises from Si donors while the higher one was
associated with a second donor of unknown origin, presumably O.\@ As we
show in Table I, Si has an EMT ionization energy of 30.4 meV with a
$1s-2p_{o}$ transition (which does not shift appreciably with field) at
22.8 meV.\@ These results would indicate that the binding energy
attributed in Ref.\ [\ref{Moore}] to Si is probably due to O, and that
the second donor may be indeed due to residual C-impurities. Since
residual C was not detected, and the samples were growth by HPVE with
no carbon sources, it would seem unlikely that carbon donors would be
present. However, as we mentioned in the introduction, small C
concentrations of $\le 2\times 10^{16}$ cm$^{-3}$ have been found in
unintentionally doped samples grown by HPVE.\cite{Zhang} Those
concentrations were below the detection limit of the secondary ion-mass
spectroscopy (SIMS), but well within the photoluminescence
limits.\cite{Zhang} This would indicate that similar SIMS non-detection
of C impurities in samples obtained by other growth techniques does not
unequivocally prove the absence of small quantities of this donor
species, allowing our identification to be correct.  Carefully
monitored C doping together with infrared absorption measurements would
be needed to clarify this point.

 We would like to point out that the transition $1s-2p_{\pm}$ we find
in Si at 22.3 meV, is in good agreement with the transition at 21.7 meV
observed by Wang {\it et al.}\cite{Wang} by infrared absorption
spectroscopy of purposely Si-doped GaN grown by HVPE.  Since the
linewidths observed were rather large ($\sim 5$ meV) compared with
widths reported for nominally undoped samples, \cite{Meyer} it is
likely that the small difference of 0.6 meV here is within the
uncertainty of the peak position measured.

Notice also that due to the anisotropy of the effective mass and
dielectric constant in the wurtzite phase of GaN, the $2p_{o}$ and
$2p_{\pm}$ states are non-degenerate even at zero field. This effect is
exhibited in Figure 1 as a shift on the transition energy of about 0.5
meV for both impurities. Notice that in Ref.\ [\ref{Moore}] a
transition at about 17 meV has also been observed, which remained
unaffected as the magnetic field was turned on. That feature was
assigned to the $1s-2p_{o}$ transition of the donor with 31.1 meV
binding energy, arguing that its origin was related to the
non-isotropic conduction band of GaN. As we can observe in Table I and
Figure 1, our EMT calculations indicate that such kind of transition
does not exist in either donor considered here.  Since the associated
binding energy of such state would be much smaller, the identity of
that species is not any of the obviously present, and it remains a
puzzle.

In Figure 2 we show the transition energy for O and C donors as a
function of the magnetic field for c-GaN. As seen there and in Table I,
the ionization levels for this phase are close to those of the
hexagonal-GaN. We find that the difference between the two phases is
typically 1 meV for each donor, being larger in the hexagonal-GaN host,
partly due the larger effective mass. Notice also that the $1s-2p_{o}$
and $1s-2p_{\pm}$ transitions are all degenerate at zero magnetic
field, as one expects in this isotropic-band system. Unfortunately, no
definitive experimental results exist yet for the shallow donor levels
in c-GaN with which to compare our results. \cite{De-Sheng}

Finally, we should mention that in all our EMT calculations we have
employed (were available) the experimental values of the effective mass
and dielectric constant for GaN. We find that donor levels are somewhat
sensitive to the variation of these parameters. For example, an
uncertainty of $\pm$ 0.01$m_o$\, \cite{Perlin1} in the conduction
effective mass would yield a change of less than $\pm$ 1.5 meV on the
energy levels.  Notice, however, that all energy levels and transitions
shown in Table I would shift correspondingly, maintaining the relative
position for the different impurities, which helps in the determination
of their identity (e.g., largest binding energy for C).

To conclude, let us also mention that we find that Ge donors show a
spectrum very similar to that found in Si. \\

We thank D. Drabold for very helpful discussions on the subtleties of
pseudopotentials.  This work was supported in part by grants ONR-URISP
N00014-96-1-0782, DURIP N00014-97-1-0315, and from CONACyT-M\'{e}xico.

\clearpage
\begin{table}
\begin{center}
\begin{tabular}{lcccccccc}
& \multicolumn{3}{c}{c-GaN}
& \multicolumn{5}{c}{h-GaN} \\
\hline
               &       &      &       &       &       &       &        &\\
               &  Si   &  O   &   C   &       &  Si   &   O   &   C    &\\
               &       &      &       &       &       &       &        &\\
\hline
$ 1s $    & 29.5  & 30.4 & 32.5  &       & 30.4  &  31.4 &  33.7  &\\
$1s-2p_{\pm}$  & 21.8  & 22.7 & 24.4  &       & 22.3  &  23.3 &  25.1  &\\
$1s-2p_{o}$    & 21.8  & 22.7 & 24.4  &       & 22.8  &  23.7 &  25.6  &\\
$1s-3p_{o}$    & 26.1  & 26.9 & 29.0  &       & 27.0  &  28.0 &  30.1  &\\
\end{tabular}
\end{center}

\caption{Ground state $1s$ and lower few internal transitions for
different impurities in GaN for both crystalline phases, zincblende
(cubic), and wurtzite (hexagonal). Magnetic field is zero here. All
energies are in meV.}

\end{table}

\begin{figure}[hb]
\epsfysize=3.in
\epsfxsize=3.5in
\epsfbox{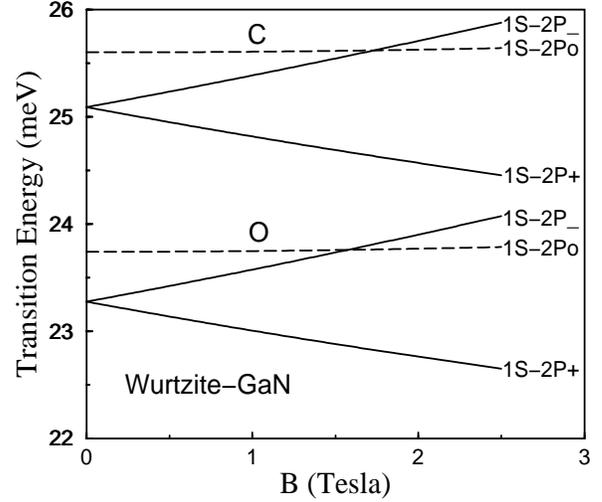}

 \caption{Zeeman splitting for O and C shallow donors in
 hexagonal GaN. Magnetic field is applied parallel to the
 c-axis.  Anisotropy in the effective mass leads to a
 shift of the $1s-2p_{o}$ transition line at $B=0$, while
 transitions $1s-2p_{\pm}$ are degenerate.}

\end{figure}

\begin{figure}[hb]
\epsfysize=3.in
\epsfxsize=3.5in
\epsfbox{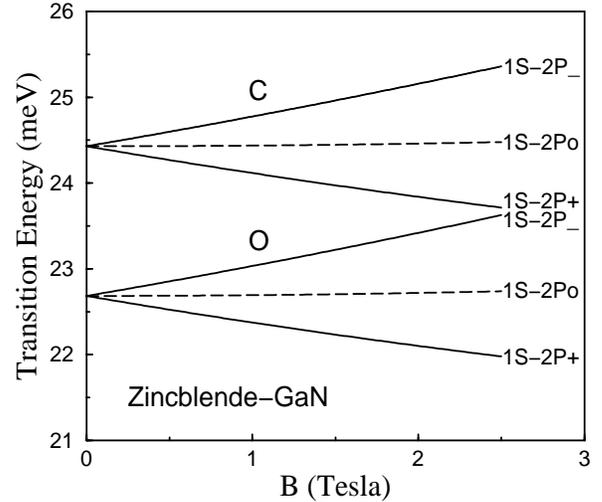}

 \caption{Zeeman splitting for O and C shallow donors in cubic
GaN.  Magnetic field is applied parallel to a symmetry axis.
Isotropy of the effective mass in this case yields the same value for
the $1s-2p_{o}$ and $1s-2p_{\pm}$ transitions at $B=0$. At $B \neq 0$
the transitions $1s-2p_{\pm}$ split accordingly.}

 \end{figure}


\begin{references}
\bibitem{spin} Since in GaN the Lande factor for the conduction band
  is $g \sim 1.95$ (E.R. Glaser {\em et al}., Phys. Rev. B. {\bf 51}, 13
  326 (1995)), and for magnetic fields up to $B=2.5$T, the energy
  associated with the interaction of the spin with the magnetic field
  is $\leq 0.15$ meV, which corresponds to a shift in the energies
  levels of $\leq$ 0.5\%.
\end{references}
\end{document}